\documentclass[10pt,letterpaper]{article}
\usepackage[top=0.85in,left=2.75in,footskip=0.75in]{geometry}

\usepackage{amsmath,amssymb}

\usepackage{changepage}

\usepackage[utf8x]{inputenc}

\usepackage{textcomp,marvosym}

\usepackage{cite}

\usepackage{nameref}

\usepackage[hidelinks]{hyperref}

\usepackage{microtype}
\DisableLigatures[f]{encoding = *, family = * }

\usepackage[table]{xcolor}

\usepackage{array}

\newcolumntype{+}{!{\vrule width 2pt}}

\newlength\savedwidth

\raggedright
\usepackage[aboveskip=1pt,labelfont=bf,labelsep=period,justification=raggedright,singlelinecheck=off]{caption}

\newcommand{\mvector}[1]{\mathbf{\MakeLowercase{#1}}}

\usepackage{amsthm}
\theoremstyle{definition}

\newtheorem{obj}{Objective}
\newtheorem{hyp}{Hypothesis}

\newcommand{\NaApp}{{\textrm{Na}^+}}
\newcommand{\NaAv}{{\textrm{Na}^-}}

\usepackage{fancyvrb}

\DefineVerbatimEnvironment{Highlighting}{Verbatim}{commandchars=\\\{\},fontsize=\small}

\makeatletter
\renewcommand{\@biblabel}[1]{\quad#1.}
\makeatother

\usepackage{lastpage,fancyhdr,graphicx}
\usepackage{epstopdf}
\fancyheadoffset[L]{2.25in}
\fancyfootoffset[L]{2.25in}
\begin{document}
\vspace*{0.2in}

\begin{flushleft}
{\Large
\textbf\newline{A multiple attribute model resolves a conflict between additive and multiplicative models of incentive salience} 
}
\newline
\\
Benjamin J. Smith\textsuperscript{1,2*},
Stephen J. Read\textsuperscript{1}
\\
\bigskip
\textbf{1} Department of Psychology, University of Southern California, Los Angeles, CA, United States of America.
\\
\textbf{2} Department of Psychology, University of California at Los Angeles, Los Angeles, CA, United States of America.
\\
\bigskip

%
%





* benjsmith@gmail.com

\end{flushleft}
\section*{Abstract}

A model of incentive salience as a function of stimulus value and interoceptive state has been previously proposed \cite{zhang2009neural}. In that model, the function differs depending on whether the stimulus is appetitive or aversive; it is multiplicative for appetitive stimuli and additive for aversive stimuli. The authors argued it was necessary to capture data on how extreme changes in salt appetite could move evaluation of an extreme salt solution from negative to positive. We demonstrate that arbitrarily varying this function is unnecessary, and that a multiplicative function is sufficient if one assumes the incentive salience function for an incentive (such as salt) is comprised of multiple stimulus features and multiple interoceptive signals.  We show that it is also unnecessary  considering the dual-structure approach-aversive nature of the reward system, which results in separate weighting of appetitive and aversive stimulus features.



\section*{Introduction}

Understanding the mechanisms involved in learning to respond to rewards and punishments (reward learning) is a central question in psychology, neuroscience, economics, and related fields.  One major distinction that researchers have made is between \textit{model-based} and \textit{model-free} responding \cite{glascher2010states,doya2002multiple}.  \textit{Model-based responding} and prediction assumes that the organism learns and uses mental representations or models of the environment to make predictions and decisions. In contrast, \textit{model-free responding} assumes that the organism relies on cached estimates of the value of actions and that its predictions are based on associative processes.   These two types of responding have very different implications for the nature of the representations and the computational processes that characterize these two forms of responding.

A further classical distinction is between \textit{instrumental} and \textit{Pavlovian} responding. In \textit{instrumental responding}, an organism learns to make a response to achieve a goal (e.g., get food), whereas in \textit{Pavlovian responding} the organism acts based on learned associations between the presence of a stimulus in the environment (CS) and the occurrence of a reward (or punishment) (UCR). Researchers note that instrumental responding can be either model-based or model-free.  According to a model-based view of instrumental learning and responding, the organism learns a model of the environment and then when deciding how to respond, reasons over that model of the world before choosing a response. Whereas according to a model-free view of instrumental learning and responding, the organism learns associations between rewards and instrumental responses and responds automatically to the appearance of a reward.

In contrast, \cite{dayan2014model} note that the typical view of Pavlovian responding is that it is solely model-free, relying on cached estimates of the value of stimuli, and not on a model of the environment.  However, they argue that recent research (see \cite{robinson2013instant,tindell2009dynamic}) provides compelling evidence that Pavlovian predictions can rely on a form of model-based prediction, as the organism can update its predictions and judgments on the fly, based on its current interoceptive state.  This work shows that changes in interoceptive state can lead to changes in the incentive value of a Pavlovian CS, without the necessity of the organism actually re-experiencing the reward (UCR), suggesting the use of some form of internal model. 

They note \cite{dayan2014model} that while there is extensive work on computational accounts of model-based instrumental responding \cite{daw2005uncertainty}, an adequate computational account of model-based Pavlovian responding is currently lacking. They note one possible mathematical model \cite{zhang2009neural} of such Pavlovian predictions and its sensitivity to interoceptive state. Along with the authors of that model, they admit that the model is somewhat unsatisfactory and they express the need for continuing work toward a more adequate model.  Here we present a more principled mathematical model.  

\subsection*{Model-based responding and revaluation}

The phenomenon of re-valuation (henceforth \textit{revaluation}) is taken as a key test of whether an animal’s behavior is model-free or model-based.  If an animal can change its evaluation of the attractiveness of a reward, without having to explicitly re-experience the reward, then this is taken as evidence that they have an explicit representation of the reward and decision-making is at least partially model-based.  For example, consider an instrumental learning task, where a rat learns to press a lever for food: If the animal is subsequently fed to satiety, and when presented with the lever now presses it less vigorously, this revaluation is taken as evidence that the response is model-based and not model-free.  The idea is that when the animal’s internal state (hunger) changes, they can revalue the attractiveness of the food without having to actually re taste the food reward.  This is taken as evidence that the animal has a sensory representation of the food and can use that to re evaluate the attractiveness of the food, thus the response is model-based.  On the other hand, if the animal’s lever pressing is unaffected by the level of hunger of the animal, this is taken as evidence of model-free responding.  

Although most researchers have considered Pavlovian responding to be solely model-free, Berridge and his colleagues \cite{robinson2013instant, tindell2009dynamic} provide compelling evidence that Pavlovian responding can be model-based.  They showed experimentally, in several tests of Pavlovian conditioning and responding, that changes in interoceptive or bodily state can shift the incentive salience (motivational value) for a CS from negative to positive, without the organism having an opportunity to explicitly re-experience the UCR.  This strongly argues that the animal had an internal representation of the UCR and was able to use that to re-evaluate the attractiveness of the UCR.

 A 2009 study \cite{tindell2009dynamic} examined the impact of changes in interoceptive state for salt appetite on incentive salience. Experimenters measured ventral pallidum firing rates  (encoding reward value), when a rat was exposed to a CS that had been associated with an extremely salty solution (hence the term Dead Sea Salt study). Firing rates were compared across several successive conditions. First, rats in a normal sodium state underwent Pavlovian conditioning.  In one condition rats learned to associate a CS with a very salty solution (UCR) (Dead Sea Salt).  In another condition they learned to associate a different CS with a sucrose solution.  In a third condition, the CS didn’t predict anything.  
 During training, only the CS for sucrose elicited high levels of firing. On a following day a state of sodium depletion was induced to create a salt appetite.  Then the different CSs were presented multiple times in extinction (i.e., no presentation of the UCR). At that point, from the very first presentation of the CS (i.e., without any re-tasting),  the salt CS elicited a level of firing as high or higher than the sucrose CS.  This suggests that changes in interoceptive state (the induced salt appetite) led to large changes in the incentive salience of the CS associated with the extremely salty solution. 

In a subsequent study \cite{robinson2013instant}, changes in interoceptive state for the extremely salty solution also led to strong changes in approach versus aversive behavior to a lever associated with the salt solution, as well as changes in gene expression related to the Fos protein in reward related regions.  In this experiment rats learned that different levers (CSs) were associated with either extreme saltiness or sweetness.  On some trials, a lever accompanied by a distinctive sound was inserted through the wall of the chamber, accompanied by a squirt of a disgusting salt solution.  On other trials, on the opposite side of the chamber, a lever was inserted with a different distinctive sound accompanied by a squirt of sweet sucrose solution.  The rats approached and nibbled the sucrose associated lever, but actively avoided the extreme salt solution associated lever. On a subsequent day the rats were injected with two drugs that mimicked the effects of two hormones that are normally associated with salt appetite (which they had never previously encountered).  Now the salt associated lever became highly attractive and received approach activity close to the level of the sucrose lever, whose attractiveness was unaffected. Further, expression of Fos proteins in reward related regions of the brain (core and rostral shell of the nucleus accumbens, as well as the ventral tegmental area (VTA) and the rostral ventral pallidum and infralimbic and orbitofrontal regions of the prefrontal cortex) was “dramatically upregulated by the combination of (a) reencountering the CS+ lever simultaneously with (b) being in the new appetite state” \cite{robinson2013instant}.

These two studies \cite{robinson2013instant, tindell2009dynamic} taken together provide strong evidence that the rats were able to re evaluate the CS, without having to re experience the UCR. This strongly argues that the rats had a representation or model of the UCR and that this model guided their response to the CS, supporting the idea that Pavlovian learning can be model-based.  

\cite{zhang2009neural} sought to develop a mathematical model of the effects of interoceptive state on incentive salience in these experiments. Berridge \cite{robinson2013instant,zhang2009neural,berridge2009dissecting} had originally argued that the impact of interoceptive state on Wanting could be captured by a multiplicative function of interoceptive state and cue strength.  This is the incentive salience or motivational value of a stimulus. 

A simple multiplicative equation in \cite{zhang2009neural} describes incentive salience or motivational value of the current CS in terms of $\kappa$, representing the level of bodily state, and $r$, representing reward value:

\begin{equation}\label{eq:zhang2009_multiplicative}
    \tilde{r}(r_t,\kappa)=\kappa r_t
\end{equation}

This multiplicative model captured the pattern of results in situations in which revaluation of the moderate solution results in a shift from somewhat positive (or neutral) to more positive or \textit{vice versa}.  However, the same shift did not occur for the Dead Sea Salt solution, in which revaluation leads to a shift from highly negative to somewhat positive.  Applying a multiplicative model in the extremely salty, Dead Sea Salt condition predicted that changes in interoceptive state would lead to a reversal in preference ordering: rats would move from preferring the higher-salt, Dead Sea Salt solution to the moderate-salt solution. However, the data showed a maintenance in preference ordering as the valuation shifted with interoceptive state. Contrary to the preference predicted in Eq.~\ref{eq:zhang2009_multiplicative}, in a salt-depleted condition, rats still preferred the moderate-salt solution to the strong-salt solution.  To capture the pattern for aversive cues, they suggested an additive function that added the log of $\kappa$, which maintains the preference ordering:  

\begin{equation}
    \tilde{r}(r_t,\kappa)= r_t + \text{log} \kappa
\end{equation}

Thus, in their account \cite{zhang2009neural} shifting among positive evaluations is modeled by a multiplicative function, whereas shifting from negative to positive is modeled by an additive function.  They and \cite{dayan2014model} implied that this was somewhat inelegant. However, neither \cite{zhang2009neural} nor \cite{dayan2014model} provided an alternative model. The problem of accounting for all the data here in a single model we describe as the non-monotonicity problem.

\subsection*{A parsimonious multi-attribute model}

Here we propose a single mathematical model, using just a multiplicative function, that can be used to capture both types of revaluation and solve the non-monotonicity problem.  There are four key hypotheses  of the models presented, outline below. These are simulated retasting, multiple attributes to a stimulus, potentially separable interoceptive responses that are relevant to different attributes, and a dual-track appetitive-aversive system. However, for our proposal below to explain the \cite{zhang2009neural} data, one must only accept Hypothesis~\ref{hyp:simulatedretasting} and either Hypothesis~\ref{hyp:multipleattributes}-\ref{hyp:multipleinteroceptivestates} \textit{or} Hypothesis~\ref{hyp:twosystems}. Hypothesis~\ref{hyp:multipleattributes}-\ref{hyp:multipleinteroceptivestates} are included because they are plausible, given the multifaceted nature of biological feedback systems, and could explain a wide variety of behavior beyond the simple example presented here.  Hypothesis~\ref{hyp:twosystems} is included because it fits with an overwhelming amount of neuroanatomical data \cite{dickinson1979appetitive,lang1995emotion,seymour2005opponent}. We hold that all are likely true.

\begin{hyp}\label{hyp:simulatedretasting}
Consistent with \cite{robinson2013instant,tindell2009dynamic,zhang2009neural,dayan2014model}, after initial training the organism has associated the CS with a sensory representation of the UCR (e.g, the Dead Sea Salt solution) and when it is faced with a decision in a new interoceptive state, it can reactivate that representation and use the representation to reevaluate it. That is, in the Dead Sea Salt studies the rat can `imaginarily retaste' the solution, or simulate tasting the solution, when it receives the appropriate CS. Parallels could be drawn between this process and those outlined in the Somatic Marker Hypothesis, in which the brain learns an `as if' body loop\cite{damasio1996somatic,bechara2004role} that can simulate interoceptive signals using top-down processes.
\end{hyp}

\begin{hyp}\label{hyp:multipleattributes}
For any given stimulus, there may be multiple sensory associations, or attributes. The different attributes can be separately represented in the learned sensory representation. For example, the rat may represent a salty solution both positively as potential for fulfilling a need for salt, and negatively as potential for hypernatremia (i.e., blood salt concentration reaching unhealthily high levels). Moderate solutions may only be weakly associated with hypernatremia but strongly associated with fulfilling a need for salt, while both solutions could be strongly associated with fulfilling salt need. Alternatively
,
we might posit need for both sodium and hydration. In this case, both solutions work to fulfill sodium need, but decrease perceived hydration level. If the moderate solution is only weakly associated with a decrease in hydration, but the strong solution is strongly associated with it, then there could also be a non-linear relationship between salt content and preference. A third plausible interpretation may be an evolved instinctual aversion to strong salt taste--a disgust response--existing simultaneously with a taste (learned or innate) for salt consumption. \end{hyp}

\begin{hyp}\label{hyp:multipleinteroceptivestates}
There are potentially different interoceptive states associated with each of the stimulus attributes.  Interoceptive states will moderate the incentive salience or motivational value of attributes that are relevant to that interoceptive state, but will not impact the value of unrelated attributes.  This suggests that the overall Wanting or lack of Wanting might be the sum of a multiplicative function of current interoceptive state(s) times the separate valuation of attributes relevant to the interoceptive state.  That is, there are potentially separate terms for each interoceptive state and its relevant attributes and overall Wanting is the sum of those terms.\end{hyp} 

\begin{hyp}\label{hyp:twosystems}
Evaluation of the stimulus (e.g., the salt solution) takes place in two independent systems, an Approach or Appetitive system, and an Avoidance or Aversive system.  There is considerable evidence for these two distinct systems and for the idea that the appetitive and aversive aspects of a stimulus are evaluated independently in two separate systems \cite{dickinson1979appetitive,lang1995emotion,seymour2005opponent}.  For example, in the Dead Sea Salt studies \cite{zhang2009neural}, an appetitive desire for salt (to ensure the body doesn't reach hyponatremia, or unhealthily low salt levels) would be evaluated by the Appetitive system, while an aversive drive to avoid salt (to avoid hypernatreia)  would be evaluated in the Aversive system.  \end{hyp}

For example, assume that in the Dead Sea Salt study \cite{zhang2009neural}, the appetitive system represents the potential of each salt solution for salt satiation, while the aversive system represents its potential for salt hypernatremia. In the animal’s normal state, the interoceptive state or salt need will be modest, the resulting weight will be small, and as a result the incentive salience for the saltiness will be low.  Thus, it will have little effect on positive valuation of either the moderate or strong solution, while the extreme aversiveness of the strong solution will lead to a strong negative weight, and a strong negative valuation in the Aversive system. When the rat is salt depleted, its interoceptive state or Need for salt goes up dramatically.  The result is that the multiplicative weight on ``saltiness'' goes up strongly for the salt depleted rat and the incentive salience of the ``saltiness'' will go up dramatically.  Both solutions will become more appetitive. However, there remains an aversive association between the strong solution and hypernatremia. Thus, although the moderate solution remains attractive, the reaction to the strong solution's net value will be conflicted, more balanced by appetitive and aversive urges, and thus the strong solution will not be as appetitive as the moderate solution (in fact, the strong solution may remain aversive). As a result, the overall valuation of both salt solutions becomes more positive, but the moderate solution may remain the more appetitive one.

\section*{Model 1: Minimal}
 
Here we first present a minimal model. The minimal model has only those features necessary to demonstrate how a multi-attribute model solves the non-monotonicity problem 
in \cite{zhang2009neural}. Later, we explore how that model could be modified and extended to model separate appetitive and aversive drives.

This model is presented in its full form online (see the Supporting Information S1), along with an interactive widget allowing readers to modify code and see the model's impact with arbitrary values.

\subsection*{Model objectives}
This minimal example sets out to solve the following model objectives:
\begin{obj}\label{Obj:SaltDepletion}Under at least weak to moderate salt depletion, the rat prefers the moderate solution to the strong solution; it may approach the strong solution sometimes, but less often than the moderate solution.\end{obj}
\begin{obj}\label{Obj:SaltSatiation}
Under salt satiation, the rat prefers the moderate solution to the strong solution (i.e., finds it less aversive).
\end{obj}
\begin{obj}\label{Obj:SomeWantingForStrong}
Under sufficiently strong salt depletion, the strong solution can become appetitive while still being less appetitive than the moderate solution.
\end{obj}
\begin{obj}\label{Obj:StrongWhenNoOtherOption}
The rat's choice of the moderate or salty solution will be a function of the relative incentive salience or "wanting" of the two solutions, where incentive salience is a function of cue strength for the two solutions. Thus, under salt depletion, the rat will take the strong solution if there is no cue for the moderate solution.
\end{obj}
\begin{obj}\label{Obj:Sugar}
Appetitive states work in parallel, so a neural response to the sugar solution is unaffected by changes in interoceptive states unrelated to sugar. Thus, when there are both a salt solution and a sugar solution, and a need for salt and sugar, a region representing a wanting signal will fire for the salt solution, as well as for the sugar solution.  The two alternatives compete. When there is not a need for salt, the wanting signal for sugar will fire regardless, while the wanting signal for salt will not \cite{tindell2009dynamic}.
\end{obj}

\subsection*{Objectives~\ref{Obj:SaltDepletion} and \ref{Obj:SaltSatiation}: multiple stimulus features}

Following and extending terminology in \cite{zhang2009neural}, we can specify the basic mathematical model that satisfies the first two objectives:

\begin{equation}\label{eq:BasicSaltModelZB}
\tilde r(\mvector{r},\mvector{\kappa})=\kappa_{\text{Na}} r_{\text{Na}}+\kappa_{h} r_{h}
\end{equation}

\noindent $\kappa$ values represent interoceptive drive, while $r$ represents expected reward. Subscripts denote the particular drive (for $\kappa$) or reward (for $r$) associated with each term: $\text{Na}$ for salt, and $h$ for hydration level. The total reward associated with a particular action is the expected outcome ($r$) associated with that action across all of the drives. 

We can recast Eq.~\ref{eq:BasicSaltModelZB} as a set of $\kappa$ values and $r$ vectors representing the drive and reward values of each alternative

\begin{align}\label{eq:BasicSaltModelZBVector}
\tilde r(\mvector{r},\mvector{\kappa})=& \mvector{\kappa}_{\text{Na}} \mvector{r}_{\text{Na}}+\mvector{\kappa}_{h} \mvector{r}_{h}
\end{align}

Each $\mvector{r}$ is a column vector of values for each solution,

\begin{equation} \nonumber
    \begin{bmatrix}\text{Moderate Na solution} \\ \text{Strong Na solution}\end{bmatrix}
\end{equation}

\noindent We can then calculate the desire for salt and the predicted action under conditions of either salt deprivation ($\kappa_{Na}=1$) or salt satiation ($\kappa_{Na}=0$). 
Two need states are posited - a `salt need' and a `hypernatremia avoidance need'. We assume that a moderately salty solution isn't associated with hypernatremia, while a strongly salty solution is associated with hypernatremia. The expected value for a moderately salty solution, $r_{\text{Na}}$, is 0.5 while for the strong solution it is 1.0. The expected value for hypernatremia $r_{h}$, is 0.0 for the moderately salty drink but -2.0 for the strongly salty solution. Then, when interoceptive drive state for salt is $\kappa_{\text{Na}}$ is 1.0, the moderately salty solution will be preferred. When it is $-1.0$, i.e., an aversion to salt due to oversatiation, the moderately salty solution will still be preferred, while the strongly salty solution will be even less preferred.

Thus, we can calculate the expected reward for each alternative, given present interoceptive signals, under salt deprivation:

\begin{align}
\tilde r(\mvector{r},\mvector{\kappa})= & \mvector{\kappa}_{\textrm{Na}} \mvector{r}_\textrm{Na}+\mvector{\kappa}_{h} \mvector{r}_{h}  \\ \nonumber
 = & 1 \times \begin{bmatrix}0.5 \\ 1.0\end{bmatrix} + 2 \times \begin{bmatrix}0.0 \\ -1.0\end{bmatrix} \\ \nonumber
 = & \begin{bmatrix}0.5 \\ -1.0\end{bmatrix} \begin{matrix} \textit{Moderate Salt solution}\\ \textit{Strong Salt solution} \end{matrix}
\end{align}

\noindent indicating a drive for the moderate solution but not the strong solution. We can combine this with a softmax function \cite{sutton1998reinforcement}
\begin{align}
P_t(a) = \frac{\exp(\tilde r_a/\tau)}{\sum_{i=1}^n\exp(\tilde r_t(i)/\tau)} \text{,}
\end{align}

\noindent in order to make the decision probabilistic. This way, either action has some probability of being performed but the higher the $\tilde r$ value, the higher probability the action will be performed. Thus, in this instance, there is a higher probability of drinking the moderate salt solution than the strong salt solution, but both choices are possible; the relative likelihood is set by the temperature $\tau$.

Under salt over-satiation:

\begin{align}
\tilde r(\mvector{r},\mvector{\kappa})= & \mvector{\kappa}_{\textrm{Na}} \mvector{r}_\textrm{Na}+\mvector{\kappa}_{h} \mvector{r}_{h}  \\ \nonumber
 = & -1 \times \begin{bmatrix}0.5 \\ 1.0\end{bmatrix} + 2 \times \begin{bmatrix}0.0 \\ -1.0\end{bmatrix} \\ \nonumber
 = & \begin{bmatrix}-0.5 \\ -3.0\end{bmatrix} \begin{matrix} \textit{Moderate Salt solution}\\ \textit{Strong Salt solution} \end{matrix}
\end{align}

\noindent indicating an aversive drive away from the moderate solution, but an even stronger drive away from the strong solution.

Under the conditions assumed here, we meet Objectives 1-2 above. 

\subsection*{Objective~\ref{Obj:SomeWantingForStrong}}

In the previous section we demonstrated meeting Objectives \ref{Obj:SaltDepletion}-\ref{Obj:SaltSatiation}. But notice that the expected value estimated for the strong solution is always negative, and we can expect the rat will not drink the salt solution, even if it is the only solution available. To meet Objective~\ref{Obj:SomeWantingForStrong} above--to show the strong salt solution can become weakly appetitive under the right conditions--we first need to assume a different weighting of hypernatremia to deprivation. For instance, we can do that by increasing the deprivation/interoceptive state value $\kappa_{\text{Na}}$ from 1 to 3. 

\subsubsection*{Results}

In that greater deprived state,

\begin{align}
\tilde r(\mvector{r},\mvector{\kappa})= & \mvector{\kappa}_{\textrm{Na}} \mvector{r}_\textrm{Na}+\mvector{\kappa}_{h} \mvector{r}_{h}  \\ 
 = & 3 \times \begin{bmatrix}0.5  \\ 1.0\end{bmatrix} + 2 \times \begin{bmatrix}0.0 \\ -1.0\end{bmatrix} \nonumber \\
 = & \begin{bmatrix}1.5 \\ 1.0\end{bmatrix} \begin{matrix} \textit{Moderate Salt solution}\\ \textit{Strong Salt solution} \end{matrix} \nonumber
\end{align}

\noindent and the strong solution now has positive value, though still less than the moderate solution. Note that with a sufficiently strong weighting towards salt need, the model would actually predict a reference for the stronger solution; this is discussed further below. The preference for the moderate solution is still greater than that for the strong solution. Importantly, if there was no moderate solution then the positive expected value for the strong solution would lead to drinking the strong solution.

\subsection*{Objective~\ref{Obj:StrongWhenNoOtherOption}: Environmental cues}

Now, in order to make explicit the role of environmental cues in generating  incentive salience we need to introduce an extra vector describing environmental cues that cue reward objects, $\mvector{c}$. Following the S-R-O expectancy valence account given in \cite{zhang2009neural}, the extra parameter describes environmental features eliciting a response. 

In this model, a response is only taken if the environment elicits it. `Elicitation' refers to a cue prompting a response by an organism
. We extend \cite{zhang2009neural}, in which the product of the environmental cue strength and interoceptive urge is a measure of the strength of incentive salience, `Wanting', or reward prediction value. The multiple factors introduced in the previous section allow us to consistently express this as a multiplicative function rather than switch between multiplicative and additive functions, as in \cite{zhang2009neural}.

Internal cues could also potentially be represented using $\mvector{c}$, although we have not needed to model that process for our current purposes. For instance, processes such as default-mode network mind-wandering, or other top-down processes including controlled processing, may draw attention to a particular stimulus, and combined with an extant need, may thereby prompt an organism to seek the reward.


\subsubsection*{Results}

Using the new model with our vector representing environmental cues $\mvector{c}$, we can predict behavior where there is no cue for a moderate salt solution\footnote{Note the use of elementwise multiplication between the matrices using the symbol $\circ$}:

\begin{align}\label{eq:WithCues}
\tilde r(\mvector{r},\mvector{\kappa})= & \mvector{\kappa}_{\textrm{Na}} \mvector{r}_\textrm{Na}\mvector{c}+\mvector{\kappa}_{h} \mvector{r}_{h}\mvector{c} \\
 = & 3 \times \begin{bmatrix}0.5 \\ 1.0\end{bmatrix}\circ \begin{bmatrix}0.0 \\ 1.0\end{bmatrix} + 2 \times \begin{bmatrix}0.0 \\ -1.0\end{bmatrix}\circ \begin{bmatrix}0.0 \\ 1.0\end{bmatrix} \nonumber \\
 = & \begin{bmatrix}0.0 \\ 1.0\end{bmatrix} \begin{matrix} \textit{Moderate Salt solution}\\ \textit{Strong Salt solution} \end{matrix} \nonumber
\end{align}

The result is that the expected value for the strong solution is 1.0 while the expected value for the moderate solution is 0.0 (because it simply hasn't been cued or elicited). The strong solution becomes the preferred solution, without salt depletion increasing, thanks to the lack of availability of the moderate solution as an alternative.

Cue should generally have a lower-bound of 0, and and for some applications it may be suitable to give cue an upper-bound of 1. For the current application, setting a particular upper-bound is not particularly important.

Cue $\mvector{c}$ models the mental accessibility of the stimulus as a whole rather than the accessibility of particular stimulus aspects. Therefore, it is also normally expected to be equal across all needs modeled. Thus, $\mvector{c}$ appears without any subscript in Eq.~\ref{eq:WithCues}.

\subsection*{Objective~\ref{Obj:Sugar}: Sugar}

The model in Eq.~\ref{eq:BasicSaltModelZBVector} can be extended slightly to account for the evidence discussed in the sugar example in \cite{tindell2009dynamic}, where the neural response to a sugar solution was unaffected by changes in interoceptive states unrelated to the sugar:

\begin{equation}\label{eq:SaltSugarModelTindell}
\tilde r(\mvector{r},\mvector{\kappa})=\kappa_{\text{Na}} r_{\text{Na}}+\kappa_{h} r_{h} + \kappa_{\text{Glc}} r_{\text{Glc}}
\end{equation}

With this simple model, we can account for the effect described in \cite{zhang2009neural} and in \cite{tindell2009dynamic}.  


\subsubsection*{Results}

If $\tilde r(\mvector{r},\mvector{\kappa})$ represents reward value, then we can very simply show how this might work using our new Eq.~\ref{eq:SaltSugarModelTindell}:

\begin{align}
\begin{split}\label{eq:SaltSugarModelTindell2}
\tilde r(\mvector{r},\mvector{\kappa})= & \kappa_{\text{Na}} r_{\text{Na}}+\kappa_{h} r_{h} + \kappa_{\text{Glc}} r_{\text{Glc}} \\
= & \kappa_{\text{Na}} \times \begin{bmatrix}0.5 \\ 1.0 \\ 0.0\end{bmatrix} + \kappa_{h} \times \begin{bmatrix}0.0 \\ -1.0 \\ 0.0\end{bmatrix} + 1 \times \begin{bmatrix}0.0 \\ 0.0 \\ 1.0\end{bmatrix} \\
= & \begin{bmatrix}0.5\kappa_{\text{Na}} \\ 1.0\kappa_{\text{Na}} \\ 0.0\end{bmatrix} + \begin{bmatrix}0.0 \\ -1.0\kappa_{h} \\ 0.0\end{bmatrix} + \begin{bmatrix}0.0 \\ 0.0 \\ 1.0\end{bmatrix} \\
 = & \begin{bmatrix}0.5\kappa_{\text{Na}} \\ 1.0\kappa_{\text{Na}}-1.0\kappa_{h} \\ 1.0\end{bmatrix} \begin{matrix} \textit{Moderate Salt solution}\\ \textit{Strong Salt solution} \\ \textit{Sugar solution}\end{matrix}
 \end{split}
\end{align}

In this model, while the reward value for the moderate and strong salty solutions will vary according to the current interoceptive craving for salt ($\kappa_{\text{Na}}$) and hydration ($\kappa_{h}$), the desire for glucose will remain constant so long as the interoceptive craving for sugar does not change.

\subsection*{Combined model}

Finally, we assume that cue strength will be a relevant factor for the sugar reward seeking, relative to salt reward seeking, so we extend the model further by adding a cue strength multiplier as described in Eq.~\ref{eq:WithCues}:

\begin{align}
\begin{split}\label{eq:SaltSugarModelWithCue}
\tilde r(\mvector{r},\mvector{\kappa})= & \kappa_{\text{Na}} r_{\text{Na}}\mvector{c}+\kappa_{h} r_{h}\mvector{c} + \kappa_{\text{Glc}} r_{\text{Glc}}\mvector{c} 
\end{split}
\end{align}

\section*{Model 2: Appetitive and aversive drives}

There is overwhelming evidence for separate neurological pathways for appetitive and aversive drives, and as a result, appetitive and aversive cues are separately evaluated in the two systems \cite{dickinson1979appetitive,lang1995emotion,seymour2005opponent}.  Here we present a model that explicitly represents those two different systems. 

\subsection*{Design}
We can modify Eq.~\ref{eq:WithCues} to calculate separate appetitive and aversive drives. To model Objectives 1-4 (the salt models, though not the sugar model) this time, we assume the same salt need and hypernatremia avoidance drive, but model the hypernatremia pathway as inherently aversive. We use the subscripts $\NaApp$ and $\NaAv$ to distinguish sodium need (the appetitive drive) and hypernatremia danger (the aversive drive) in these instances.

\begin{align}\label{eq:AppetitiveAversive1}
\tilde r(\mvector{r},\mvector{\kappa})= & \mvector{\kappa}_{\textrm{Na}^+} \mvector{r}_{\NaApp}\mvector{c}-\lambda(\mvector{\kappa}_{\NaAv} \mvector{r}_{\NaAv}\mvector{c})
\end{align}

Here, $\lambda$ represents a loss aversion parameter and describes how aversive losses are compared to gains. From a purely behavioral modeling perspective, $\lambda$ is not necessary because it simply represents a weight to the general interoceptive effect of aversive $\mvector{r}$ relative to appetitive $\mvector{r}$. However, theoretical and empirical research has established the phenomenon of \textit{loss aversion} \cite{tversky1991loss,tom2007neural}, in which a loss is more aversive than an equally sized gain.  Neurobiologically, there appears to be separable circuitry for reward and punishment \cite{frank2005dynamic,cox2015striatal}. Specifically, excitatory, approach action is driven by a direct, striatal dopaminergic pathway, while aversive action is driven through other means, including an indirect inhibitory dopaminergic pathway. Thus, we can expect the relative strength of those two systems could affect the relative strength of all aversive and appetitive attributes in a model. Thus, a value like $\lambda$ which captures that relative strength is conceptually useful when linking the model in Eq.~\ref{eq:AppetitiveAversive1} to its neurobiological correlates.

\subsection*{Results}

The next few equations work out how each of the four objectives above can be described using this model. To meet Objective~\ref{Obj:SaltDepletion}, we assume that the moderate salt solution is preferred to the strong solution because hypernatremia risk is lower, even though it provides less salt satiation. We assume loss aversion $\lambda=2$:

\begin{align}\label{eq:AppetitiveAversive2}
\tilde r(\mvector{r},\mvector{\kappa})= & \mvector{\kappa}_{\textrm{Na}^+} \mvector{r}_{\NaApp}\mvector{c}-\lambda(\mvector{\kappa}_{\NaAv} \mvector{r}_{\NaAv}\mvector{c}) \\
 = & 1 \times \begin{bmatrix}1.0 \\ 2.0\end{bmatrix}\circ \begin{bmatrix}1.0 \\ 1.0\end{bmatrix} - 2\times2 \times \begin{bmatrix}0.1 \\ 1.0\end{bmatrix}\circ \begin{bmatrix}1.0 \\ 1.0\end{bmatrix} \nonumber \\
 = & \begin{bmatrix}0.6 \\ -2.0\end{bmatrix} \begin{matrix} \textit{Moderate Salt solution}\\ \textit{Strong Salt solution} \end{matrix} \nonumber
\end{align}

We can show that the rat will naturally also prefer the moderate solution under salt satiation (Objective~\ref{Obj:SaltSatiation}), though both are at least slightly aversive:

\begin{align}\label{eq:AppetitiveAversive3}
\tilde r(\mvector{r},\mvector{\kappa})= & \mvector{\kappa}_{\textrm{Na}^+} \mvector{r}_{\NaApp}\mvector{c}-\lambda(\mvector{\kappa}_{\NaAv} \mvector{r}_{\NaAv}\mvector{c}) \\ \nonumber
 = & 0.1 \times \begin{bmatrix}1.0 \\ 2.0\end{bmatrix}\circ \begin{bmatrix}1.0 \\ 1.0\end{bmatrix} - 2\times 2 \times \begin{bmatrix}0.1 \\ 1.0\end{bmatrix}\circ \begin{bmatrix}1.0 \\ 1.0\end{bmatrix} \\ \nonumber 
 = & \begin{bmatrix}0.1 \\ 0.2\end{bmatrix} - \begin{bmatrix}0.4 \\ 4.0\end{bmatrix} \\ \nonumber
 = & \begin{bmatrix}-0.3 \\ -3.8\end{bmatrix} 
 \begin{matrix} \textit{Moderate Salt solution}\\ \textit{Strong Salt solution} \end{matrix}
 \nonumber
\end{align}

Further, under sufficiently strong salt depletion there may be \textbf{some} desire for the stronger solution (Objective \ref{Obj:SomeWantingForStrong}):

\begin{align}\label{eq:AppetitiveAversive4}
\tilde r(\mvector{r},\mvector{\kappa})= & \mvector{\kappa}_{\textrm{Na}^+} \mvector{r}_{\NaApp}\mvector{c}-\lambda(\mvector{\kappa}_{\NaAv} \mvector{r}_{\NaAv}\mvector{c}) \\
 = & 3 \times \begin{bmatrix}1.0 \\ 2.0\end{bmatrix}\circ \begin{bmatrix}1.0 \\ 1.0\end{bmatrix} - 2\times 2 \times \begin{bmatrix}0.1 \\ 1.0\end{bmatrix}\circ \begin{bmatrix}1.0 \\ 1.0\end{bmatrix} \nonumber  \\ 
 = & \begin{bmatrix}3 \\6\end{bmatrix} - \begin{bmatrix}0.4 \\ 4\end{bmatrix} \nonumber \\ 
 = & \begin{bmatrix}2.6 \\ 2\end{bmatrix} 
 \begin{matrix} \textit{Moderate Salt solution}\\ \textit{Strong Salt solution} \end{matrix}
 \nonumber
\end{align}

Finally, we can show that, under salt depletion, the rat will take the strong salt solution if it has no other option (Objective~\ref{Obj:StrongWhenNoOtherOption}):

\begin{align}\label{eq:AppetitiveAversive5}
\tilde r(\mvector{r},\mvector{\kappa})= & \mvector{\kappa}_{\textrm{Na}^+} \mvector{r}_{\NaApp}\mvector{c}-\lambda(\mvector{\kappa}_{\NaAv} \mvector{r}_{\NaAv}\mvector{c}) \\
 = & 3 \times \begin{bmatrix}1.0 \\ 2.0\end{bmatrix}\circ \begin{bmatrix}0.0 \\ 1.0\end{bmatrix} - 2\times2 \times \begin{bmatrix}0.1 \\ 1.0\end{bmatrix}\circ \begin{bmatrix}0.0 \\ 1.0\end{bmatrix} \nonumber \\
 = & \begin{bmatrix}0 \\6\end{bmatrix} - \begin{bmatrix}0 \\ 4\end{bmatrix} \nonumber \\ 
 = & \begin{bmatrix}0 \\ 4\end{bmatrix} 
 \begin{matrix} \textit{Moderate Salt solution}\\ \textit{Strong Salt solution} \end{matrix}
 \nonumber
\end{align}

\section*{Discussion}

Our model is inspired by the possibility that in the normal ecology of an organism, such as a rat or a human, there will be multiple possible stimuli to which the organism can respond (Hypothesis~\ref{hyp:multipleattributes}) (e.g, water, food, conspecific, possible mate) and there will be related multiple potential interoceptive states (Hypothesis~\ref{hyp:multipleinteroceptivestates}).  In our earlier work \cite{READ2017237} (see also \cite{keramati2014homeostatic,keramati2017misdeed,revelle2015model}) we have argued that decision-making will be the result of a competitive process among currently activated motives, where the strength of the motive activation or Wanting is a multiplicative function of the relevant interoceptive state and the strength of the relevant cue.  Thus, any decision or choice is going to depend on the relative (and not the absolute) strength of alternative motives.  Moreover, this strongly suggests that in real ecologies one needs to consider all the relevant possibilities and cannot focus on just one stimulus or motive.

A key assumption in our model, following \cite{dayan2014model,zhang2009neural}, is that during Pavlovian learning the organism develops a sensory representation of the reward and that it can use that representation to simulate the "retasting" or re experiencing of the reward.  Given that assumption there are two separate mutually consistent pathways proposed here by which the data regarding salt revaluation \cite{zhang2009neural,tindell2009dynamic} may be explained. The first is that described in the previous paragraph: multiple stimulus attributes are combined with multiple interoceptive states to produce a more complex overall evaluation of a Pavlovian stimulus. The second is that appetitive and aversive stimuli are processed separately, and so when a stimulus contains both appetitive and aversive elements (e.g., a salt solution), the overall balance or predicted reward or incentive salience may depend on the strength of the relative appetitive and aversive elements. We have described both here as each independently can explain the salt revaluation data.  However, we argue that both  are actually utilized by the brain for Pavlovian responses to stimuli in the real world.

\subsection*{Limitations}



The models presented in this article describe how, even under a salt deprivation condition, an organism will prefer a moderately salty to a strongly salty solution. However, for all the models presented, once Salt Need is sufficiently strong, the strong solution will be preferred to the moderate one, a preference reversal not apparent in the observed data. We believe the model is not necessarily weakened by this possibility, because there are undoubtedly physiological upper limits to the values that salt need might take: the values that would lead to this preference reversal are likely outside the bounds of a viable biological system. The model still demonstrates parsimoniously how a moderately salty solution may be preferred to a strongly salty solution, under both aversive and appetitive states.

\subsection*{Links to other work}

The model presented here helps to tie together a number of separate lines of research. Aside from \cite{READ2017237, keramati2014homeostatic,keramati2017misdeed}, the Cues-Tendencies-Action personality model \cite{revelle2015model} is another that exemplifies how competing cues and drives can inspire different actions. The exploration of model-based Pavlovian associations has enabled us to draw links in the current article from personality-based models \cite{revelle2015model,READ2017237}, designed primarily to model model-based, instrumental activity, to models of Pavlovian activity such as the one presented in this article.

This function $\tilde r(\mvector{r},\mvector{\kappa})$ could be a candidate for an explicit mathematical definition for the drive function $\textrm{d}(h)$ in \cite{keramati2014homeostatic, keramati2017misdeed}. $\textrm{d}(h)$ in \cite{keramati2014homeostatic} describes the relative drive associated with a stimulus, and similarly to our presentation, can describe multiple drives and their interactions with one another, based on their distance in multi-dimensional space from a neutral set-point. It is roughly analogous to $\kappa$ in this paper.  However,  \cite{keramati2014homeostatic,keramati2017misdeed} have not explained how their drive framework can be used to calculate a value for $\textrm{d}(h)$ or $\tilde r$ (which has a similar meaning in their framework as here) and thus have not specified how it provides a solution for the non-monotonicity problem \cite{zhang2009neural}. The model proposed in this paper demonstrates how that link might work by describing how the  the multiplicative cue-interoception function described by \cite{zhang2009neural} might produce that drive. Our model is neutral on whether the function measures distance from a homeostatic `set point' or whether it measures allostasis and a distance from an allostatic equilibrium. In this way, the model here may be regarded as an integrative proposal combining key elements of \cite{keramati2014homeostatic} and \cite{zhang2009neural} to parsimoniously address the additive/multiplicative dilemma in \cite{zhang2009neural}.

The OpAL model \cite{collins2014opponent} also goes some of the way toward addressing the non-monotonicity problem \cite{zhang2009neural, dayan2014model}.  The OpAL model argues that in making choices an organism will separately weight the evaluations in two systems, a Go and a NoGo system, similarly to the `Model 2' Appetitive and Aversive system that we describe here.  This is consistent with a key assumption we make.  However, their weighting parameters, $\beta_G$ and $\beta_N$, describe overall DA level, not specific interoceptive states. $\beta_G$ and $\beta_N$ cannot vary in response to salt depletion and salt satiation in particular without affecting other states, but would yield a re-weighting of all positive (and negative) attributes. In contrast, prior research \cite{tindell2009dynamic} suggests  modification of salt appetite has had no affect on rats' sugar consumption-related behavior.  Thus, the OpAL model \cite{collins2014opponent} is not on its own a plausible solution in the context of a model-based Pavlovian learning system such as the one posited in \cite{zhang2009neural}. In contrast, we propose that interoceptive state, which can be specific to a particular attribute of a reward, can lead to attribute specific re-weighting. Thus, our approach argues that re evaluative re-weighting will be specific to the current interoceptive state or Need of the organism (e.g., hunger versus thirst versus affiliation), whereas the OpAL model would suggest it is globally moderated by tonic dopamine levels.  

\subsection*{Future directions}

The models presented in the present work demonstrate how the salt revaluation data \cite{zhang2009neural} may be explained using a multi-attribute Pavlovian valuation model. Such a model can of course explain a much broader array of phenomena--any instance in which a Pavlovian stimulus may contain both appetitive and aversive associations, or even any instance where there are multiple stimulus attributes that vary at least somewhat independently. A simple extension might parallel our previous work \cite{READ2017237}, demonstrating how the model extends to an arbitrary set of approach and avoid motivations. 

We have also not described the process of reward and punishment update learning in the multi-attribute context. However, this would be a relatively straightforward extension from the model presented here.

\subsection*{Conclusion}

As previously noted \cite{dayan2014model}, a variety of studies have shown that, contrary to earlier assumptions that Pavlovian responding was solely model-free, that Pavlovian responding can often be model-based, based on the simulated re evaluation of a representation of a potential reward (or threat).  Several of the most compelling examples of this process were the series of studies we described involving responses to extremely strong salt solutions \cite{zhang2009neural,tindell2009dynamic,robinson2013instant}.

Unfortunately, existing computational accounts of this process were not really adequate. For example, the previous mathematical account of this process by Zhang and his colleagues involved a multiplicative function for positive evaluation and an additive function for negative evaluation.  Here we have shown how this process can be explained more parsimoniously with a single multiplicative function. This is done by positing either (a) multiple stimulus attributes of a reward (or punishment) and multiple interoceptive responses relevant to different attributes, or (b) separate appetitive and aversive motivational pathways. It is likely that both sets of assumptions are true. The present model provides a more adequate account of Pavlovian, model-based evaluations than has been provided by previous work \cite{zhang2009neural}.

\section*{Supporting information}





\paragraph*{S1 Supplementary Online Materials.}\label{S1_SupplementaryOnlineMaterials}
{\bf Viewing and running the model online.} At the time of publication, the models described here, including python code to generate them, can be viewed in a Juypter Notebook posted on the github repository website at \url{https://github.com/bjsmith/motivation-simulation/blob/master/minimal-example.ipynb}. From that page, readers can launch an interactive tool using the python implementation to explore the model's behavior with any arbitrary set of values.

Readers can also download the model, with the demo page, from the github repository to examine or use in their own environment.


\section*{Acknowledgments}
We are indebted to John R. Monterosso and Brandon Turner for their pre-reviews and helpful suggestions for this paper.

\section*{Funding}

Research reported in this publication was supported by the National Institute Of General Medical Sciences of the National Institutes of Health under Award Number R01GM109996. The content is solely the responsibility of the authors and does not necessarily represent the official views of the National Institutes of Health. 


%

%
%

\bibliography{biblio}

\end{document}